\documentclass[preprint]{aastex6}
\usepackage{graphicx}
\usepackage{amssymb}
\usepackage{amsmath}
\usepackage{wasysym}

\begin{document}

\title{Bayes' theorem and early solar short-lived radionuclides: the case for an unexceptional origin for the solar system}

\author{Edward D. Young}
\email[]{eyoung@epss.ucla.edu}
\affil{Department of Earth, Planetary, and Space Sciences, University of California Los Angeles, 595 Charles E. Young Drive East, 
Geology Building, Los Angeles, CA 90095-1567, USA}


\date{\today}

\begin{abstract}
The presence of excesses of  short-lived radionuclides in the early solar system evidenced in meteorites has been taken as testament to close encounters with exotic nucleosynthetic sources, including supernovae or AGB stars.  An analysis of the likelihoods associated with different sources of these extinct nuclides in the early solar system indicates that rather than being exotic, their abundances were typical of star-forming regions like those observed today in the Galaxy.  The radiochemistry of the early solar system is therefore unexceptional, being the consequence of extensive averaging of molecular cloud solids.   
\end{abstract}

\keywords{Solar system: abundances: radionuclides }
\maketitle

\section{INTRODUCTION}

The conventional view of the origin of our solar system has been that the abundances of the radionuclides, especially the short-lived radionuclides (SLRs) with mean lives of several million years or less, are testament to close encounters with particular, often exotic, nucleosynthetic sources, including collapse supernovae (SNe) or AGB stars \citep{Camer77,Wasse96,Wasse2006,Lugar2014}. Indeed,  apparent excesses in SLRs have even been used as evidence for a supernova trigger for the formation of the solar system, an idea that has waxed and waned but persists still \citep{Camer77a,Goune2012}.  Here it is shown that the initial abundances of the SLRs in the early solar system are likely typical of star-forming regions rather than exotic.  This result stands in contrast to the notion that the solar-system SLRs were overabundant relative to normal molecular cloud material and instead indicates that the radiochemistry of the early solar system is typical of massive star-forming regions (SFRs) like those observed today in the Galaxy; a comprehensive theory for the radiochemistry of the solar system is best cast in terms of a statistical analysis of SFR material rather than close encounters with a finite number of individual nucleosynthesis sources. The success of this approach in explaining the relative abundances of solar-system radionuclides places the origin of the solar system squarely in the realm of business as usual for SFRs and suggests that the abundances of individual nuclides should not be taken as evidence for individual stellar sources.

Theories for the provenance of the solar-system radionuclides can be divided into two types of scenarios. In the first scenario type, the radiochemistry of the solar system has no significance beyond chance encounters between pre-solar materials and a variety of nucleosynthesis sources \citep{Wasse96,Wasse2006}. In the second scenario type, the birth environment of the solar system was like the self-enriched massive star-forming regions (SFRs) of today \citep{Jura2013}, and the abundances of the radionuclides are the result of extensive averaging in the SFR.  Some models are a mix of the two types \citep[e.g.,][]{Goune2012}.  The two scenario types are described in \S 2 and \S 3 below and their relative likelihoods are compared in \S 4.  Section 5 describes implications for the nature of the short-lived radionuclide carriers.   

\section{CHANCE ENCOUNTERS}
The chance-encounter scenarios emphasize the discrete nature, or ``granularity'' \citep{Wasse96}, of stellar nucleosynthesis events that could have seeded parental solar system material with nuclides.  They can be described by an equation based on a geometric series summing individual nucleosynthesis events (encounters) with an average temporal spacing $\delta t$ followed by a final actual free decay time $\Delta t$  \citep{Wasse2006,Lugar2014} 

\begin{equation}
\frac{{N_{\text{R}} }}
{{N_{\text{S}} }} = \left[ {\frac{{P_{\text{R}} }}
{{P_{\text{S}} }}\;\frac{\delta t }
{T}\left( {1 + \frac{{\operatorname{e} ^{ - \delta t /\tau } }}
{{1 - \operatorname{e} ^{ - \delta t /\tau } }}} \right)} \right]\exp ( - \Delta t/\tau )
\end{equation}

\noindent where $N_i$ and $P_i$ are number and production rates for radionuclides (R) and stable isotope partner (S), respectively, $T$ is the age of the Galaxy (7.4 Gyr at the time of solar system formation), $\Delta t$ is in this case the time interval between the last event (LE) and the birth of the solar system, and $\tau$ is the mean life of R against radioactive decay. Dividing $N_{\rm R}$ by $N_{\rm S}$  eliminates the effects of element-specific chemistry when evaluating Equation (1) for different radionuclides.  $N_{\rm R}/N_{\rm S}$ ratios are well known for the early solar system from precise measurements of the radioactive decay products of the SLRs and the relative abundances of the longer-lived nuclides in meteoritical materials \citep{Huss2009}.  Comparisons of different radionuclides are further facilitated by dividing the abundance ratio on the left-hand side of Equation (1) by the production ratio on the right-hand side of Equation (1).  Production ratios are obtained from models of stellar nucleosynthesis \citep{Rausc2002,Woosl2007}. The resulting quotient, $ (N_{\rm R}/N_{\rm S})/(P_{\rm R}/P_{\rm S})$, for different radionuclides depends on the radioactive mean lifetime $\tau$, all else equal.  Therefore, if the radionuclides inherited by the solar system from its parent molecular cloud were all of similar average age, $ (N_{\rm R}/N_{\rm S})/(P_{\rm R}/P_{\rm S})$ should vary only with $\tau$, with the shorter-lived nuclides being less abundant relative to their production rates than the longer-lived nuclides.  Equation (1) allows for differences in the ages of the nuclides depending upon the astrophysical environment in which they formed.

\begin{figure}
\centering
\includegraphics[scale=0.73]{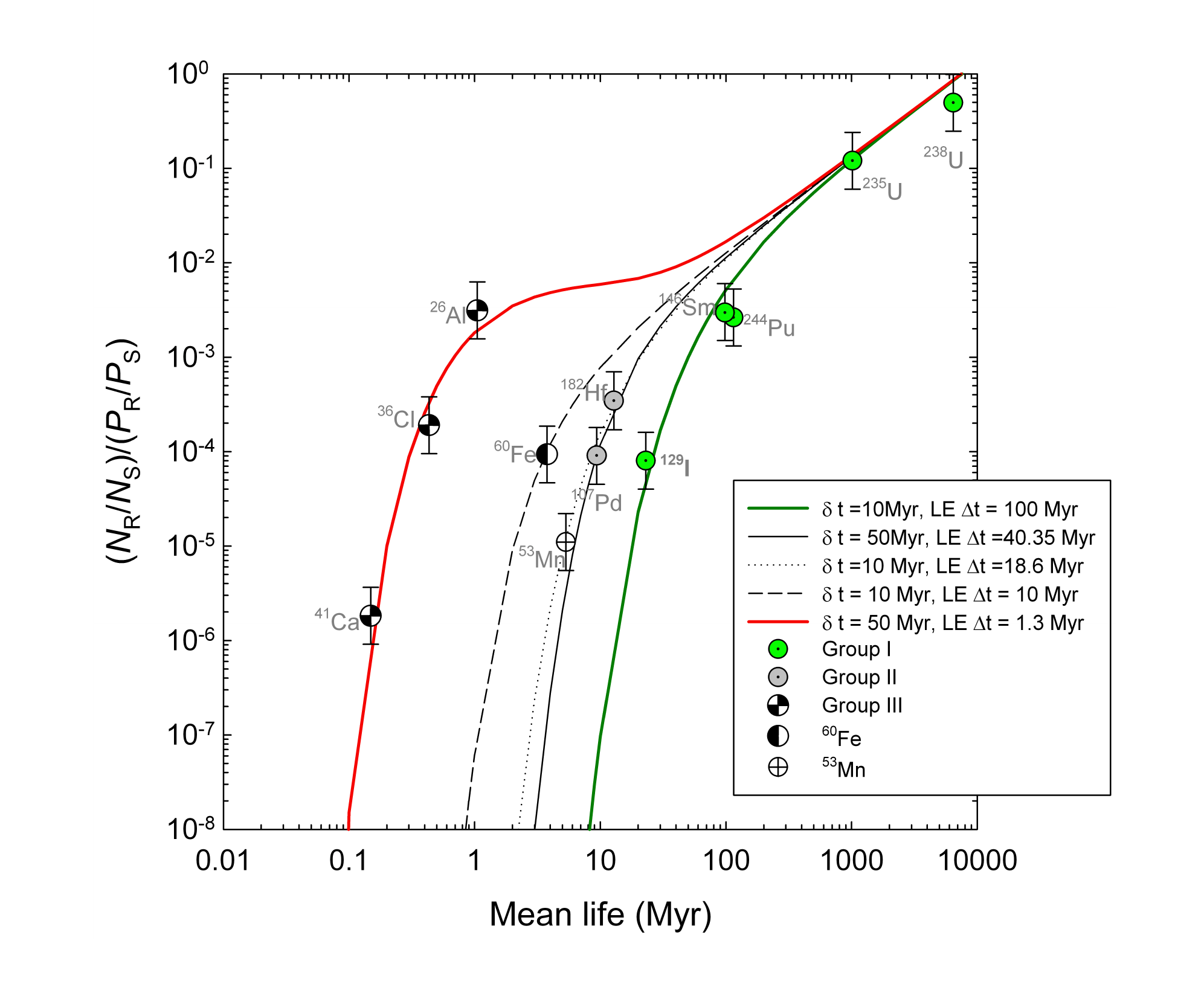}
\caption{Plot of radionuclide abundances ratioed to their stable nuclide partners ($N_{\text{R}}/N_{\text{S}}$) and production ratios ($P_{\text{R}}/P_{\text{S}}$) vs. mean decay lifetime for radionuclides present at the birth of the solar system.  The values for the abundance ratios, production ratios, and decay lifetimes are from published values \citep{Young2014} updated with new production ratios for $\rm ^{182}Hf$ and $\rm ^{107}Pd$ \citep{Lugar2014}. Model curves are from Equation (1) fit to the different nuclide groups as indicated in the legend and in the text.}
\label{Figure1}
\end{figure}

Equation (1) can be used to derive models of $\log[(N_{\rm R}/N_{\rm S})/(P_{\rm R}/P_{\rm S})]$ versus $\log({\tau})$, resulting in several groupings of radionuclides for which production ratios are well known (Figure 1, where radionuclides made primarily by spallation are omitted).  These groupings succinctly summarize many of the relationships previously described in the literature \citep[e.g.,][]{Wasse2006}.  For example, the relative concentrations of r- and (in one case) p-process products $\rm^{129}I$, $\rm ^{146}Sm$, $\rm ^{244}Pu$, $\rm ^{235}U$ and $\rm ^{238}U$ are all explained by Equation (1) using $\delta t= 10$ Myr and $\Delta t = 100$ Myr (Figure 1).  The value for $\delta t$ is consistent with the frequency of supernova events affecting random positions in the Galactic disk \citep{Meyer2000} and so is appropriate for these SNe-derived nuclides. Here we label these isotopes as Group I.  Similarly, it was shown recently \citep{Lugar2014} that a single set of values for $\delta t$  and $\Delta t$ can explain the relative concentrations of both $\rm ^{107}Pd$ and $\rm ^{182}Hf$, both nuclides being dominantly s-process products from AGB stars (Figure 1).  A value for $\delta t$ of 50 Myr, appropriate for AGB star encounters, and a similar value for $\Delta t$ of 40.35 Myr fits the  $\log[(N_{\rm R}/N_{\rm S})/(P_{\rm R}/P_{\rm S})]$ values for $\rm ^{107}Pd$ and $\rm ^{182}Hf$ (Figure 1). We refer to these radionuclides as Group II.  $\rm ^{53}Mn$ and $\rm ^{60}Fe$ are treated separately from Group I and II. $\rm ^{53}Mn$  is also fit by the Group II curve but its origin must be distinct as it is a SN product, suggesting a shorter $\delta t$  interval.  $\rm ^{60}Fe$ requires its own $\Delta t$ (Figure 1). The shortest-lived nuclides, labeled Group III, have  $\log[(N_{\rm R}/N_{\rm S})/(P_{\rm R}/P_{\rm S})]$ values that are explained with Equation (1) using $\delta t= 50$ Myr and $\Delta t = 1.3$ Myr (Figure 1).  The latter model reflects the fact that these short-lived nuclides have no â``memory'' of events prior to their most recent synthesis.  Five distinct models defined by five $\Delta t$ values ($\delta t$ values are prescribed {\it a priori} by astrophysical constraints) represented by five curves are therefore required to explain the radionuclide abundances in Figure 1, one each for Groups I, II, and III and two others for $\rm ^{53}Mn$ and $\rm ^{60}Fe$. 

\section{SELF ENRICHMENT}

The self-enrichment of star-forming regions features the enhanced effects of winds from Wolf-Rayet (WR) stars.  WR stars have massive progenitors ($M_{*} > 20$ to $25 M_{\odot}$ ).  Their large progenitor masses ensure that they have short lifetimes (several Myr) and are therefore spatially correlated with star-forming regions since they don't have time to flee their birth environment before they die \citep{Young2014}.  In one recent formulation of the self-enrichment scenario \citep{Young2014}, all 12 radionuclides considered in Figure 1 are explained by a single model based on a two-phase interstellar medium (ISM)  composed of a molecular cloud phase (MC) and a diffuse phase and with a molecular cloud mass fraction $x_{\rm MC}$ of  $\sim 0.17$, equivalent to that today.  The model equation  \citep{Young2014,Jacob2005} is

\begin{equation}
  \log \left( {\frac{{{N_{{\text{R,MC}}}}}}{{{N_{{\text{S,MC}}}}}}} \right) - \log \left( {\frac{{{P_{\text{R}}}}}{{{P_{\text{S}}}}}} \right) = 2\log \tau- 
  \log \left[ {(1 - {x_{{\text{MC}}}}){\tau _{{\text{MC}}}} + \tau} \right] - \log {T} \\ 
\end{equation} 

\noindent  where the abundances of radionuclides and their stable partners now refer to those in the molecular cloud phase and environs in the SFR  ($N_{\text{R,MC}}$ and $N_{\text{S,MC}}$, respectively) as opposed to the diffuse ISM outside of the SFR.  The radionuclide production terms relevant to the molecular cloud setting are cast in terms of production from supernovae ( $P^{\text{SNe}}_{\text{R}}$) and production from WR winds ($P^{\text{W}}_{\text{R}}$): 

\begin{equation}
\frac{{{P_{\text{R}}}}}{{{P_{\text{S}}}}} = \frac{{{\Lambda _{{\text{SNe}}}}P_{\text{R}}^{{\text{SNe}}}}}{{{\Lambda _{{\text{SNe}}}}{P_{\text{S}}}}} + \frac{{{\Lambda _{\text{W}}}P_{\text{R}}^{\text{W}}}}{{{\Lambda _{{\text{SNe}}}}{P_{\text{S}}}}}
\end{equation}

\noindent where $\Lambda_{\rm W}$ and $\Lambda_{\rm SNe}$ are the relative efficiencies for trapping the two sources of nuclides in the star-forming region.  WR production values are listed by  \cite{Young2014}. The model fits the solar-system data using two independent parameters, an enhancement of WR wind production over SN production in SFRs, with $\Lambda_{\rm W}/\Lambda_{\rm SNe} \sim 4000$, and a sequestration time of nuclides in molecular cloud dust,  $\tau_{\rm MC} \sim 200$ Myr (Figure 2).   $\tau_{\rm MC}$ is the {\it average} time spent in SFR molecular clouds before exiting by dispersal or by incorporation into stars. Because individual clouds exist for shorter time spans than the SFR as a whole (see Appendix), time spent passing from one cloud to another via inter-cloud space in the SFR is included in the residence time. The radionuclide residence time $\tau_{\rm MC}$ of $\sim 200$ Myr  is consistent with the lifetime of dust in the interstellar medium \citep{Tiele2005} and the timescale for converting molecular clouds to stars \citep{Drain2011}.  In general, large values for $\Lambda_{\rm W}/\Lambda_{\rm SNe}$ are consistent with the fact that massive stars like WR progenitors apparently do not typically end their lives as energetic SNe but rather collapse by fallback to form black holes directly \citep{Fryer99,Smart2009}.  In this way, the products of WR winds are enhanced relative to SNe products in SFRs sufficiently massive as to host a population of large stars \citep{Young2014}. The fate of WR stars is debated, however, as evidenced by different interpretations of recent spectroscopy data for type IIb supernova 2013cu \citep{Galya2014,Groh2014}, making {\it a  priori} evaluation of $\Lambda_{\rm W}/\Lambda_{\rm SNe}$ tenuous.  Regardless, high values for $\Lambda_{\rm W}/\Lambda_{\rm SNe}$ on an SFR scale are also favored by the fact that most supernovae derive from stars that outlive their association with their birth environs \citep{Young2014}.   

\begin{figure}
\centering
\includegraphics[scale=0.75]{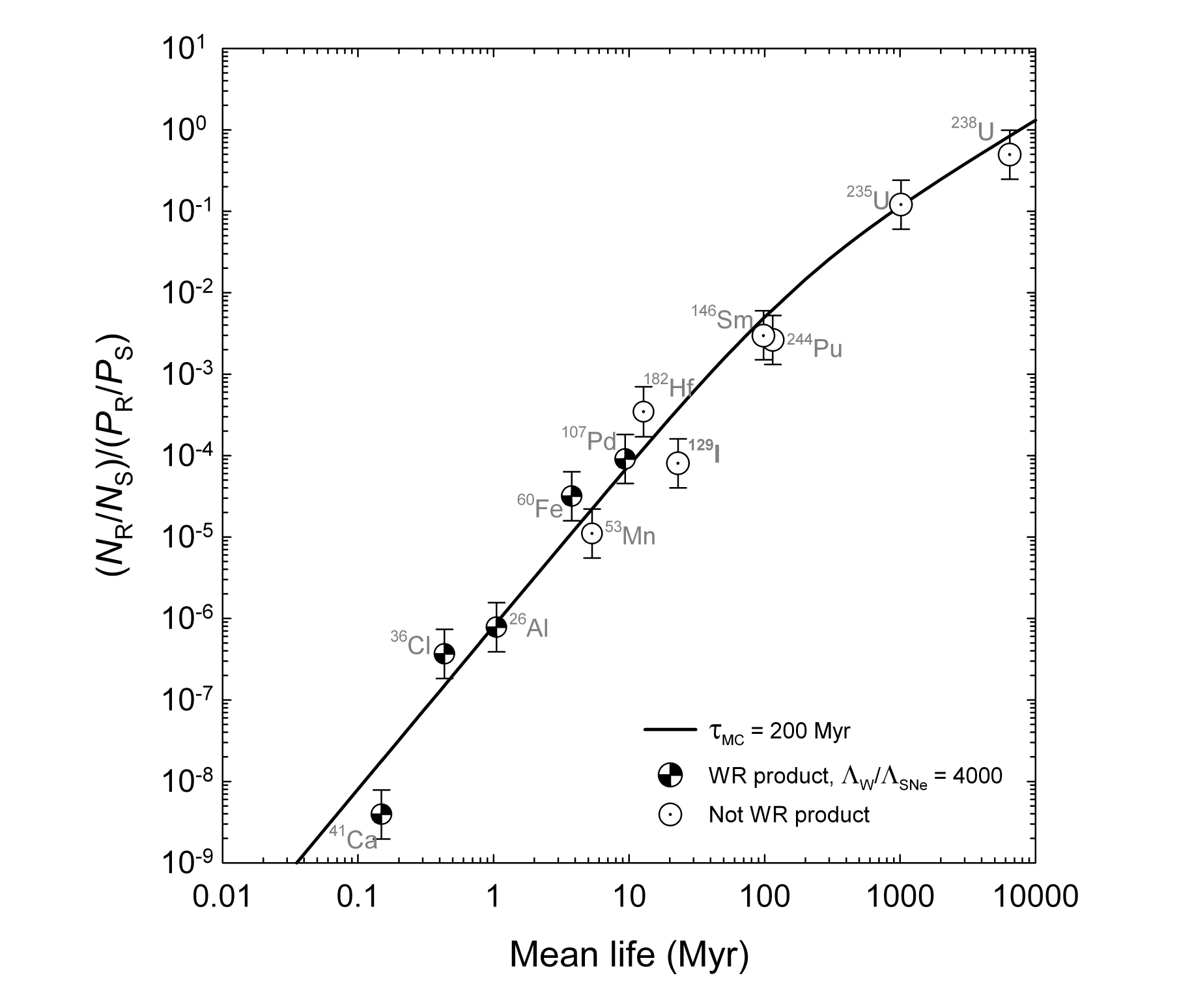}
\caption{Plot of radionuclide abundances ratioed to their stable nuclide partners ($N_{\text{R}}/N_{\text{S}}$) and production ratios ($P_{\text{R}}/P_{\text{S}}$) vs. mean decay lifetime for radionuclides in the early solar system.  The plot is similar to Figure 1 but with ordinate values calculated using production ratios that include $\Lambda_{\rm W}/\Lambda_{\rm SNe} = 4000$ and published WR production terms compiled by\cite{Young2014}.  The curve is Equation (2) fit to the data using $\Lambda_{\rm W}/\Lambda_{\rm SNe} = 4000$ and $\tau_{\rm MC} = 200$ Myr. }
\label{Figure2}
\end{figure}

\section{ASSESSING PROBABILITIES}

 In an effort to move past qualitative arguments for and against these disparate hypotheses, Bayes'/Laplace's theorem is used here  to assess the relative likelihoods for the two classes of explanations exemplified by Figures 1 and 2.  The relevant equation is \citep{Jeffr61,Schwa78}

\begin{equation}
\frac{{P(h_1 |\vec x)}}
{{P(h_2 |\vec x)}} = \underbrace {\frac{{P(\vec x|\vec \theta _1 h_1 )}}
{{P(\vec x|\vec \theta _2 h_2 )}}N^{ - 1/2(n_{\theta ,1}  - n_{\theta ,2} )} }_{{\text{Bayes Factor}}}\;\overbrace {\frac{{P(h_1 )}}
{{P(h_2 )}}}^{{\text{Priors}}}
\end{equation}

\noindent where $P(h_i |\vec x)$ is the posterior probability of hypothesis $h_i$ given data $\vec x$,  $P(\vec x|\vec \theta _i h_i)$ is the conditional probability for data $\vec x$  assuming hypothesis $h_i$ represented by parameters $\vec \theta _i$  is correct, $N$ is the number of data (in this case 12), $n_{\theta ,i}$ are the number of parameters defining the models, and $P(h_i)$ are the {\it a priori} probabilities for hypotheses $i$ independent of the data.  The conditional probabilities are assessed as the integrals of the $\chi^2$ probability densities for each fit.  The models in Figure 1 together are defined by 5 independent parameters yielding 7 degrees of freedom and a combined reduced $\chi^2$ of 1.14, corresponding to $P(\vec x|\vec \theta _1 h_1)$= 0.335 (0.5 is optimal).  The model in Figure 2 is defined by 2 independent parameters, 10 degrees of freedom, and a reduced $\chi^2$ of 0.95, corresponding to  $P(\vec x|\vec \theta _2 h_2)$ = 0.484.  Assuming equal priors of 0.5 each for now, Equation (4) leads to

\begin{equation}
\frac{{P(h_1 |\vec x)}}
{{P(h_2 |\vec x)}} = \underbrace {\frac{{0.335}}
{{0.484}}12^{ - 1/2(5 - 2)} }_{{\text{Bayes Factor}}}\;\overbrace {\frac{{0.5}}
{{0.5}}}^{{\text{Priors}}} = 0.017.
\end{equation}

\noindent This ratio, being $\ll 1$, constitutes ``strong'' evidence \citep{Kass95} that the SFR  hypothesis ($h_2$) is favored over the chance-encounter hypothesis ($h_1$).  The fits to the data are effectively equally good and can't distinguish the models.  Rather, the result in Equation (5) is attributable to the information criterion part of the Bayes Factor \citep{Schwa78} that penalizes models for many versus fewer fit parameters;  there is a quantifiable cost to probability when using more model parameters than necessary to fit the data (a quantification of Occam's razor).  

 Schwarz's formulation of the Bayesian information criterion (BIC) used as the Bayes Factor in Equation (5) is one of two common methods for assessing the relative merits of complexity versus parsimony in model selection \citep{Aho2014}.  The other is the Akaike information criterion (AIC) \citep{Aikaike1973}.  The AIC excels at selecting models with the best predictive power at the risk of unwarranted complexity while the BIC excels at selecting the``correct'' model where it is present among the models being evaluated \citep{Aho2014,Kass95}.  Because of their general nature, we assume that the two scenario types being evaluated here encompass the correct model.  In any event, application of the AIC to the analysis above gives a result similar to that in Equation (5); the probability ratio in Equation (5) is 0.05 using the AIC,   similar to the BIC-derived value of 0.02. 

The simple priors used in the analysis here are formally ``noninformative'' in so far as they imply no prior knowledge of the veracity of one hypothesis relative to the other.   An inherent lack of {\it a priori} information about various aspects of the models under consideration leads to a preference for the simple priors used here.  In principle, however, the conservative assumption of equal priors could be replaced by astrophysical constraints in the form of proper probability densities  \citep{Sivia2006}.  For example, the prior for the chance-encounter scenario could include a distribution representing the chances that an average nucleosynthesis interval $\delta t$ of 10 Myr would be followed by a significantly longer final delay $\Delta t$ of 100 Myr (i.e., many $\sigma$ from the mean interval represented by $\delta t$) as required by the $\rm ^{129}I$, $\rm ^{146}Sm$, $\rm ^{244}Pu$, $\rm ^{235}U$ and $\rm ^{238}U$ data.  Although perhaps not possible at present, a refined prior for the self-enrichment model  could depend in part on an {\it a priori} probability distribution for $\Lambda_{\rm W}/\Lambda_{\rm SNe}$.  

\section{CARRIER GRAINS}

If self-enrichment of an SFR is indeed the explanation for the solar abundances of the short-lived radionuclides, the solar-system abundances of these isotopes must be averages of progenitor dust grains with SLR abundances ranging from ~ zero (old grains) to values greater than solar (very young grains).  A model for the $\rm ^{26}Al/^{27}Al$ produced by random grain growth from large numbers of grains produced in an SFR is shown in Figure 3 following the procedures in \cite{Young2014}.  Details are presented in the Appendix. The central limit theorem ensures that the $\rm ^{26}Al/^{27}Al$ distribution of the new grains is Gaussian even though the initial distribution is heavily skewed towards very low values due to decay for $10^8$ yrs (the residence time as interstellar medium grains).  Because of the central limit theorem, the self-enrichment scenario also indicates that even small variations in isotope ratios in solar-system grains are echoes of larger (by orders of magnitude) dispersion in their molecular cloud precursors.  
 
For example, the peak in interstellar medium grain sizes is at about $0.1 \mu$m \citep{Tiele2005} while hibonite grains in meteorites, among the most primitive solar-system solids, are generally $\sim 30 \mu$m in size.  Assuming an ISM-like grain size in the star-forming clouds, it takes about 27 million cloud grains to make a single 30 $\mu$m hibonite grain (based on identical densities).  The central limit theorem states that $\sigma^\prime = \sigma(n)^{-1/2}$  where $\sigma$ is the standard deviation of the original dust, $\sigma^\prime$ is the standard deviation of the newly grown grains, and $n$ is the number of original grains averaged to make the newly grown grains.  Accordingly, the predicted variability in $\rm ^{26}Al/^{27}Al$ for the molecular cloud precursors to solar-system hibonites should be about 5000 times that exhibited by the hibonites.  Where evidence for live $\rm ^{26}Al$ is evident in hibonite grains, the initial $\rm ^{26}Al/^{27}Al$ values are about $5\times 10^{-5} \pm 1\times 10^{-5}$ \citep{Liu2008}, suggesting a range in the precursor grain  $\rm ^{26}Al/^{27}Al$ values of $\sim 0$ to $0.05$, not dissimilar to the large range in Figure 3. Similarly, it takes about 8000 hibonite grains to make the more evolved calcium-aluminum-rich inclusions (CAIs) with typical diameters of $\sim 300 \mu$m. CAIs are among the most primitive objects in carbonaceous chondrite meteorites, but are generally acknowledged to post-date the hibonite grains \citep{Liu2008}.  The dispersion about the mean in CAI initial $\rm ^{26}Al/^{27}Al$ predicted from the spread in hibonite values is $\pm 0.01\times 10^{-5}$, a value that is generally consistent with the tightly clustered CAI $\rm ^{26}Al/^{27}Al$ data \citep{Jacob2008}.  A  narrowing of the range in $\rm ^{50}Ti$ isotope anomalies of $\pm \sim 30 \permil$  in hibonites to $\pm \sim 0.1 \permil$ in CAIs is further evidence for averaging \citep{Irela88,Niede85,Chen2009}.  The reduction in dispersion from many orders of magnitude in original dust to a very narrow range of values in solar-system grains is consistent with the averaging implied by the self-enrichment origin for $\rm ^{26}Al$. 

\begin{figure}
\centering
\includegraphics[scale=0.7]{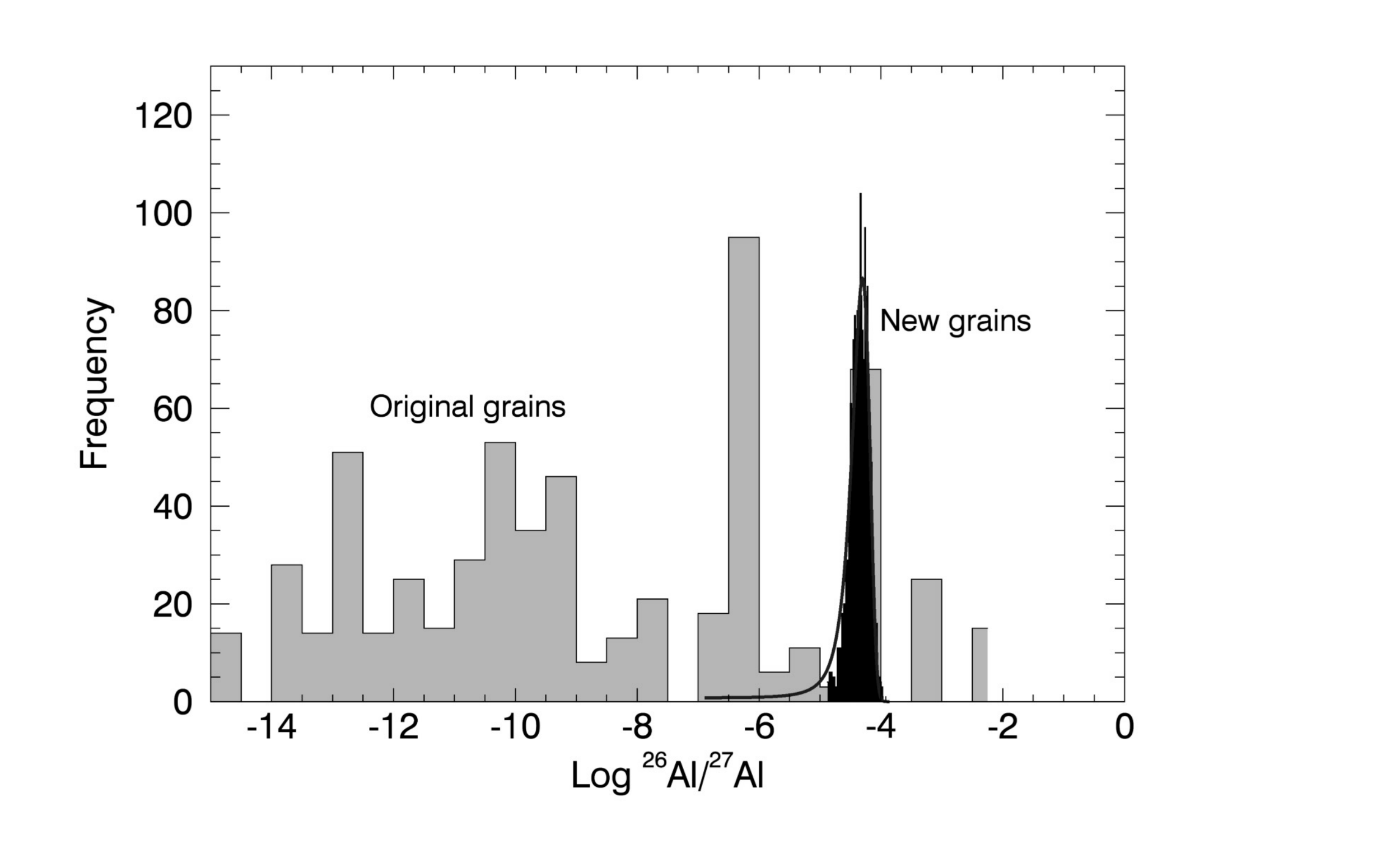}
\caption{Results of a model for $\rm ^{26}Al/^{27}Al$ hosted by grains in a star-forming region (grey) sampled at random by grain growth (black).  The resulting distribution of $\rm ^{26}Al/^{27}Al$ in the newly grown grains is Gaussian, a consequence of the central limit theorem.  The solid curve shows the Gaussian fit to the new grain distribution (note the abscissa is log scale).  The spread in original grain values over many orders of magnitude is erased by the grain growth.}
\label{Figure3}
\end{figure}

\section{CONCLUSIONS}
The inductive approach codified by the Bayes/Laplace theorem allows us to assess the relative likelihoods that the abundances of the radionuclides in the early solar system were the result of chance encounters with discrete stellar sources or instead the result of averaging over a large number of sources typical of massive star-forming regions.  The analysis favors the latter.  There is, therefore, no compelling evidence that tracing a particular nuclide to one particular stellar source (e.g., proximal supernova) is the appropriate interpretation of the radionuclide data.  Rather, the abundances of these nuclides are well explained if they were controlled by their mean lifetimes against radioactive decay and the characteristic steady-state residence time of material in star-forming regions prior to incorporation into stars.  This result suggests that the solar system formed in a large star-forming region not unlike Cygnus today.        

\acknowledgments
The author acknowledges discussions and input from Mike Jura (UCLA), Ben Zuckerman (UCLA), Ming-chang Liu (UCLA), and Matthieu Gounelle (Muséum national d'histoire naturelle, Institut universitaire de France).

\appendix
Following the procedure described by \cite{Young2014}, an IDL+Fortran code was written to simulate the production of $\rm ^{26}Al$ in the vicinity of molecular cloud material for prolonged periods of time ($\sim 100$ Myrs).  The goal is to track the availability of $\rm ^{26}Al$ to cloud material that survives multiple episodes of star formation that are spatially correlated at the kpc scale for of order $10^8$ years, as described by \cite{Elmeg2007} and \cite{Elmeg2010}.
  
Fictive star clusters were generated using the mass generation function of \cite{Kroup93} modified by \cite{Brass2006}.  With this function, initial masses for each star $j$ are drawn at random according to the expression

\begin{equation}
{M_j}/{M_\odot } = 0.01 + (0.19{x^{1.55}} + 0.05{x^{0.6}})/{(1 - x)^{0.58}}
\end{equation}

\noindent where $x$ is a uniformly distributed random number between 0 and 1.  For the results in this paper 5000 stars per cluster (i.e., 5000 random draws for $x$) were used, representing the high end of cluster populations where WR stars (i.e., O stars) are most likely to be found.  Clusters are born every $1.5$ Myr in these simulations, consistent with the cluster frequency observed in the Cygnus star forming region today \citep{Comer2012}.  The cloud material is exposed to each of the local clusters simultaneously.  In order to allow for a finite interval of star formation (i.e., all stars in a cluster are not born at exactly the same time), the birth date of each star in a cluster was altered at random over a total time interval of 1 Myr.  The exact value of this ÒblurringÓ of birth dates turns out to have little effect on the models. 

The minimum progenitor mass for WR activity is $\sim 20$ to $25 M_\odot$.  The lifetimes of stars in this mass range means that all WR stars eject winds in the vicinity of cloud material in this model.  For $5000$ stars per cluster, the average number of WR stars per cluster is $1.6 \pm 1.2$ ($1\sigma$, based on sampling 500 clusters) with a range generally between 1 and 5.  A maximum size for progenitors to form collapse supernovae was set at $20 M_\odot$ in the present calculations. 

The mass of $\rm ^{26}Al$ ejected by WR winds and supernovae proximal to cloud material is tallied as a function of time.  Integrated masses ejected by WR winds were obtained in approximate form from \cite{Goune2012}.  The total yields as a function of stellar progenitor masses used for interpolation for all masses are

\hfill \newline
\begin{center}
\begin{tabular}{ c | c }
\hline
Progenitor Mass ($M_\odot$) & $\rm ^{26}Al$ ($M_\odot$) \\
\hline
20	&   $2.0\times 10^{-10}$ \\
25	&  $2.0\times 10^{-6}$ \\
30	&  $2.0\times 10^{-5}$ \\
60	&  $7.0\times 10^{-5}$ \\
80	&  $1.0\times 10^{-4}$ \\
120	&  $2.0\times 10^{-4}$ \\
\hline
\end{tabular}
\end{center}
\hfill \newline

\noindent where the maximum stellar mass considered was $120 M_{\odot}$.  Yields for Supernovae were obtained from interpolation of values given by \cite{Chief2013}.  Exact values for yields are not important to the conclusions to be drawn from this simulation. 

For every time step each parcel of $\rm ^{26}Al$ delivered by a star that experiences winds or ends its life as a supernova within 10 Myrs (the expected time for association of a star and its natal cloud) of the initiation of the cluster was added to the total inventory of available $\rm ^{26}Al$ for the duration of the model.   For supernova ejecta a single pulse of $\rm ^{26}Al$ with mass $M_{{}^{{\text{26}}}{\text{Al,}}j}^{{\text{SN}}{{\text{e}}^{\text{o}}}}$ is added to the total inventory of $\rm ^{26}Al$ at the moment the star explodes (if the star does so within 10 Myr of the birth of its cluster).  The radioactive decay of this parcel of $\rm ^{26}Al$ from supernova $j$ is followed through time $t$ using

\begin{equation}
{M_{{}^{{\text{26}}}{\text{Al,}}j}} = M_{{}^{{\text{26}}}{\text{Al,}}j}^{{\text{SN}}{{\text{e}}^{\text{o}}}}\exp ( - t/{\tau _{{}^{{\text{26}}}{\text{Al}}}})
.
\end{equation}
                                    
Products of winds evolve in a more complicated fashion because they are delivered over extended periods of time, with time constant $\tau_{\rm W}$, simultaneous with radioactive decay with mean life $\tau_{\rm ^{26}Al}$.  Their evolution is therefore governed by the equation

\begin{equation}
\frac{{d{M_{{}^{{\text{26}}}{\text{Al}},j}}}}{{dt}} = \frac{{M_{{}^{{\text{26}}}{\text{Al}},j}^{\text{W}}}}{{{\tau _{\text{W}}}}} - \frac{{M_{{}^{{\text{26}}}{\text{Al}},j}^{}}}{{{\tau _{{}^{{\text{26}}}{\text{Al}}}}}}
\end{equation}

\noindent where $M_{{}^{{\text{26}}}{\text{Al}},j}$  is the time-dependent mass of $\rm ^{26}Al$ evolving from stellar source $j$ as modified by radioactive decay and $M^{\rm W}_{{}^{{\text{26}}}{\text{Al}},j}$ is the evolving mass of $\rm ^{26}Al$ derived from the stellar source $j$ at time $t$. The solution for each WR source $j$ is

\begin{equation}
{M_{{}^{{\text{26}}}{\text{Al}},j}}(t) = \frac{{{\tau _{{}^{{\text{26}}}{\text{Al}}}}}}{{{\tau _W} - {\tau _{{}^{{\text{26}}}{\text{Al}}}}}}M_{{}^{{\text{26}}}{\text{Al}}}^{{{\text{W}}^{\text{O}}}}(\exp ( - t/{\tau _{\text{W}}}) - \exp ( - t/{\tau _{{}^{{\text{26}}}{\text{Al}}}}))
\end{equation}

\noindent where $M_{{}^{{\text{26}}}{\text{Al}}}^{{{\text{W}}^{\text{O}}}}$  is the total integrated mass of $\rm ^{26} Al$ delivered by winds from source $j$.  Inspection of the time evolution of wind products given by \cite{Goune2012}  suggests a value for $\tau_{\rm W}$ of $\sim 8$ Myr.  Many O stars end their lives prior to this time interval, so the time evolution in Equation (3) is cut short with the death of the WR star and is replaced by simple decay of the remaining $\rm ^{26}Al$ using the form of Equation (2).  For purposes of comparison with real data, the mass of $\rm ^{27} Al$ inherited from the interstellar medium in the model was set to yield the observed solar-system $\rm ^{26}Al/^{27}Al$ for the average mass of $\rm ^{26} Al$  returned by the model.  The result for the mass of $\rm ^{26} Al$ in the SFR with respect to time is shown in Figure 1.

\begin{figure}
\centering
\includegraphics[scale=0.65]{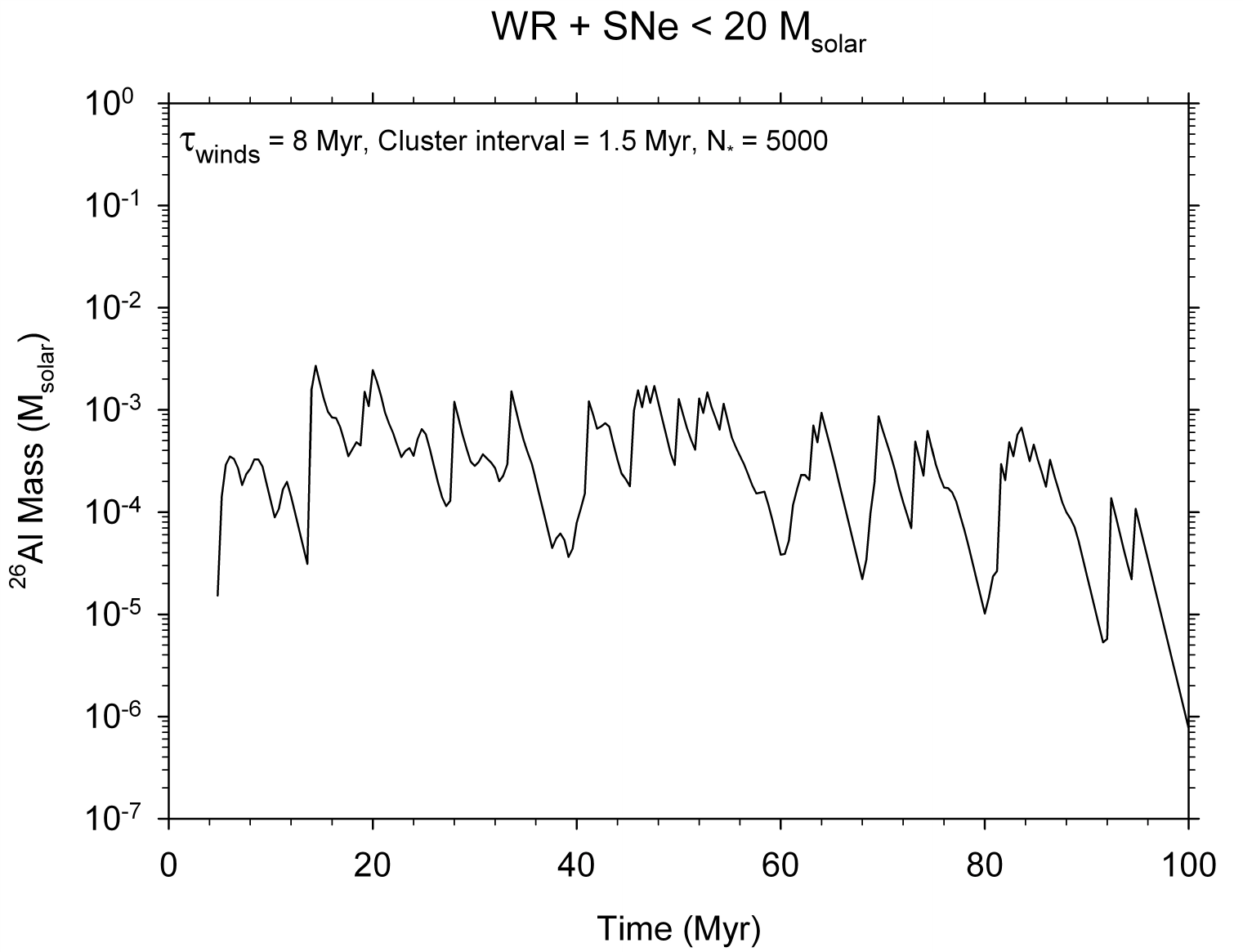}
\caption{Mass of $\rm ^{26}Al$ produced in the model described in this Appendix as a function of time in a long-lived star-forming region.  Wolf-Rayet winds dominate production of  $\rm ^{26}Al$ about 50\% of the time in these calculations.}
\label{FigA1}
\end{figure}
    
    Grain growth from the molecular cloud material was simulated by sampling and averaging the products of the evolution described above, taking into account radioactive decay of each parcel of $\rm ^{26}Al$ from the time each parcel was produced to the time of grain sampling. A grain sampling time of 100 Myr was used.   The results are not affected by the exact timing of sampling.  For example, if there are 500 bins composed of molecular cloud grains of 500 different ages, and each is sampled at random 100 aliquots at a time, and this is repeated 1000 times, the resulting distribution of 1000 newly-formed grain populations is converted from a highly asymmetrical population skewed towards effectively zero $\rm ^{26}Al$ to a Gaussian distribution.  The simulations of grain growth using this strategy are found to be in accordance with the central limit theorem such that 

\begin{equation}
\sigma^{\prime} = \sigma {(n)^{ - 1/2}}
\end{equation}
 
\noindent where $\sigma^{\prime}$ is the standard deviation for the new population of grains,  $\sigma$ is the standard deviation for the original population of smaller grains, and $n$ is the number of smaller grains comprising the larger grains.

\bibliographystyle{apj}

\end{document}